\documentclass[runningheads]{llncs}
\usepackage[T1]{fontenc}
\usepackage{amsmath}
\usepackage{graphicx}
\usepackage{multirow}
\usepackage{subfigure}
\usepackage{amssymb}
\usepackage{bm}
\usepackage[colorlinks=true]{hyperref}

\begin{document}
\title{Joint Modeling of Image and Label Statistics for Enhancing Model Generalizability of Medical Image Segmentation\thanks{Xiahai Zhuang is the corresponding author. This work was funded by the National Natural Science Foundation of China (grant no. 61971142, 62111530195 and 62011540404) and the development fund for Shanghai talents (no. 2020015).}}
\titlerunning{Deep Bayesian Segmentation}
% If the paper title is too long for the running head, you can set
% an abbreviated paper title here
%
\author{Shangqi Gao\inst{1} \and % index{Gao, Shangqi}
Hangqi Zhou\inst{1} \and % index{Zhou, Hangqi}
Yibo Gao\inst{1} \and % index{Gao, Yibo}
Xiahai Zhuang$^*$\inst{1}} % index{Zhuang, Xiahai}
% %
\authorrunning{S. Gao et al.}
% % First names are abbreviated in the running head.
% % If there are more than two authors, 'et al.' is used.
% %
\institute{School of Data Science, Fudan University, Shanghai 200433, China\\ \url{www.sdspeople.fudan.edu.cn/zhuangxiahai/}}
\maketitle              % typeset the header of the contribution
\begin{abstract} 
	Although supervised deep-learning has achieved promising performance in medical image segmentation, many methods cannot generalize well on unseen data, limiting their real-world applicability.
	To address this problem, we propose a deep learning-based Bayesian framework,
	which jointly models image and label statistics, utilizing the domain-irrelevant contour of a medical image for segmentation. 
	Specifically, we first decompose an image into components of contour and basis. Then, we model the expected label as a variable only related to the contour. Finally, we develop a variational Bayesian framework to infer the posterior distributions of these variables, including the contour, the basis, and the label. The framework is implemented with neural networks, thus is referred to as deep Bayesian segmentation.
	Results on the task of cross-sequence cardiac MRI segmentation show that our method set a new state of the art for model generalizability. Particularly, the BayeSeg model trained with LGE MRI generalized well on T2 images and outperformed other models with great margins, \textit{i.e.}, over 0.47 in terms of average Dice. Our code is available at \url{https://zmiclab.github.io/projects.html}.

\end{abstract}
\section{Introduction}

Medical image segmentation is a task of assigning specific class for each anatomical structure. Thanks to the advance of deep learning, learning-based methods achieve promising performance in medical image segmentation \cite{unet/2015,3dunet/2016,nnunet/2021}. However, many methods require a large number of images with manual labels for supervised learning \cite{zhao/2019,chen/2018}, which limits their applications. For cardiac magnetic resonance (CMR) image segmentation, repeatedly labeling multi-sequence CMR image requires more labor of experts, and therefore is expensive \cite{zhuang/2019,zhuang/2016}. Besides, the models trained at one site often cannot perform well at the other site \cite{Li/2019}. Therefore, exploring segmentation methods with better generalizability is attractive and challenging.

Much effort has been made to train an end-to-end network by supervised learning. U-Net is one of the widely used networks, since it is more suitable for image segmentation \cite{unet/2015}. Training deep neural networks in a supervised way often requires a lot of labeled data \cite{zhao/2019}, but manual labeling of medical image requires professional knowledge and is very expensive. However, small training dataset can result in the problems of over-fitting and overconfidence, which will mislead the clinical diagnosis \cite{Li/2019,ACDC/2018}. To solve the problems, Kohl et al. \cite{punet/2018} proposed a probabilistic U-Net (PU-Net) for segmentation of ambiguous images by learning the distribution of segmentation. Their results showed that PU-Net could produce the possible segmentation results as well as the frequencies of occurring. Recently, Isensee et al. \cite{nnunet/2021} developed a self-configuring method, i.e., nnU-Net, for learning-based medical image segmentation. This model could automatically configure its preprocessing, network architecture, training, and postprocessing, and achieves state-of-the-art performance on many tasks. Nevertheless, current learning-based methods deliver unsatisfactory performance when applied to unseen tasks \cite{cheng/2021}, and improving generalizability of deep learning models is very challenging.

In this work, we propose a new Bayesian segmentation (BayeSeg) framework to promote model generalizability by joint modeling of image and label statistics. To the best of our knowledge, this is the first attempt of combining image decomposition, image segmentation, and deep learning. Concretely, we first decompose an image into two parts. One is the contour of this image, and the other is the basis approximating its intensity. Both the contour and basis are unknown, and we assign hierarchical Bayesian priors to model their statistics. After that, since the contour of an image is more likely to be sequence-independent, site-independent, and even modality-independent, we try to generate a label from the contour by explicitly modeling of label statistics. Finally, given an image, we build neural networks to infer the posterior distributions of the contour, basis, and label. Being different from many deep learning models that try to learn a deterministic segmentation from a given image, BayeSeg is aimed to learn the distribution of segmentation.  

Our contributions are summarized as follows:
\begin{itemize}
	\item We propose a new Bayesian segmentation framework, i.e., BayeSeg, by joint modeling of image and label statistics. Concretely, we decompose an image into the contour and basis, and assign hierarchical Bayesian priors to model the statistics of the contour, basis, and expected label.
	\item We solve the model by developing a variational Bayesian approach of computing the posterior distributions of the contour, basis, and label, and build a deep learning architecture of implementing the approach.
	\item We validate BayeSeg on the tasks of cross-sequence segmentation and cross-site segmentation, and show the superior generalizability of BayeSeg for unseen tasks.
\end{itemize}

% %%%% Methods
\section{Methodology}
\begin{figure}[!t]
	\centering
	\subfigure[PGM]{\includegraphics[width=0.38\linewidth]{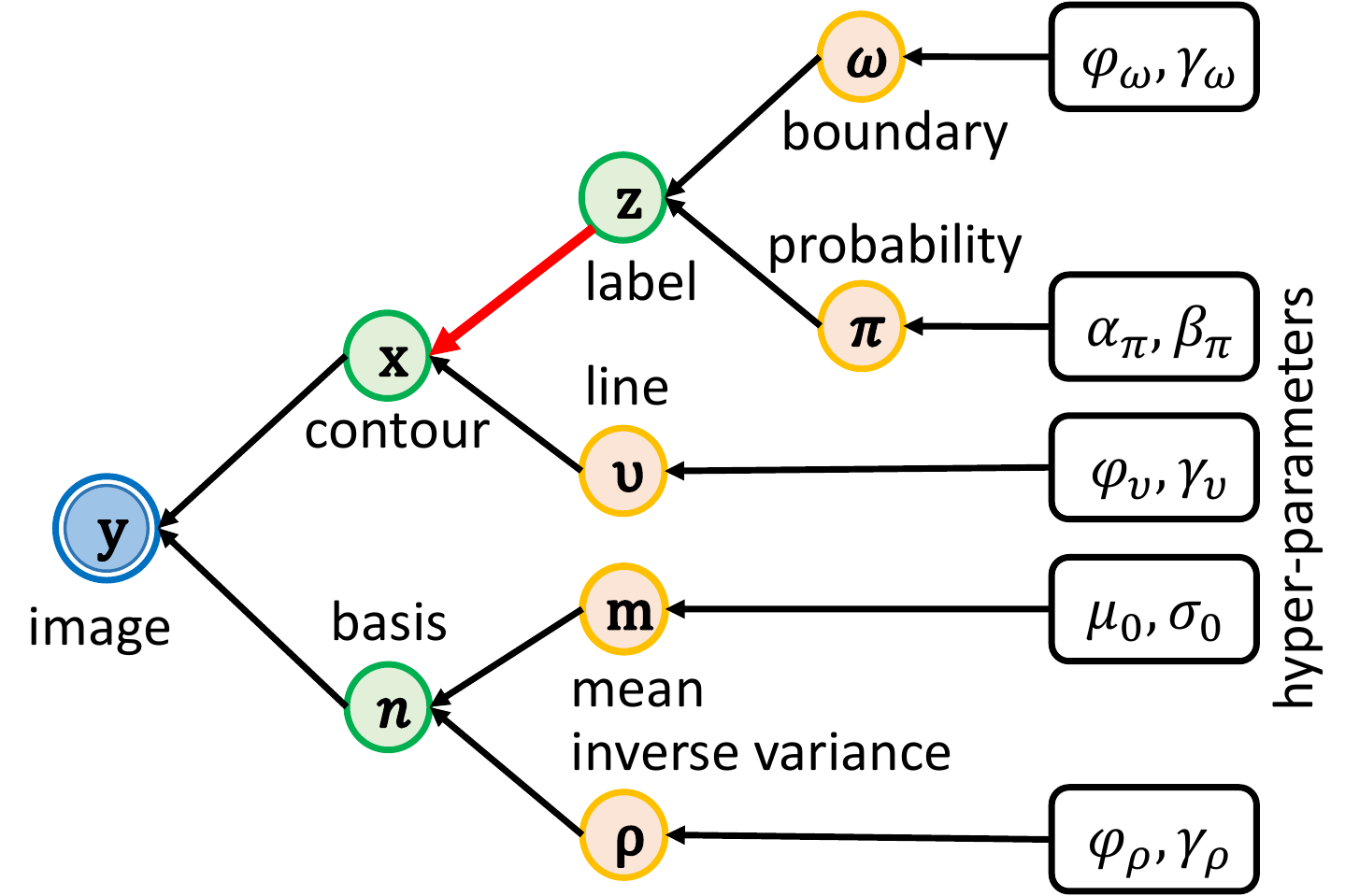}}
	\subfigure[Deep learning framework]{\includegraphics[width=0.6\linewidth]{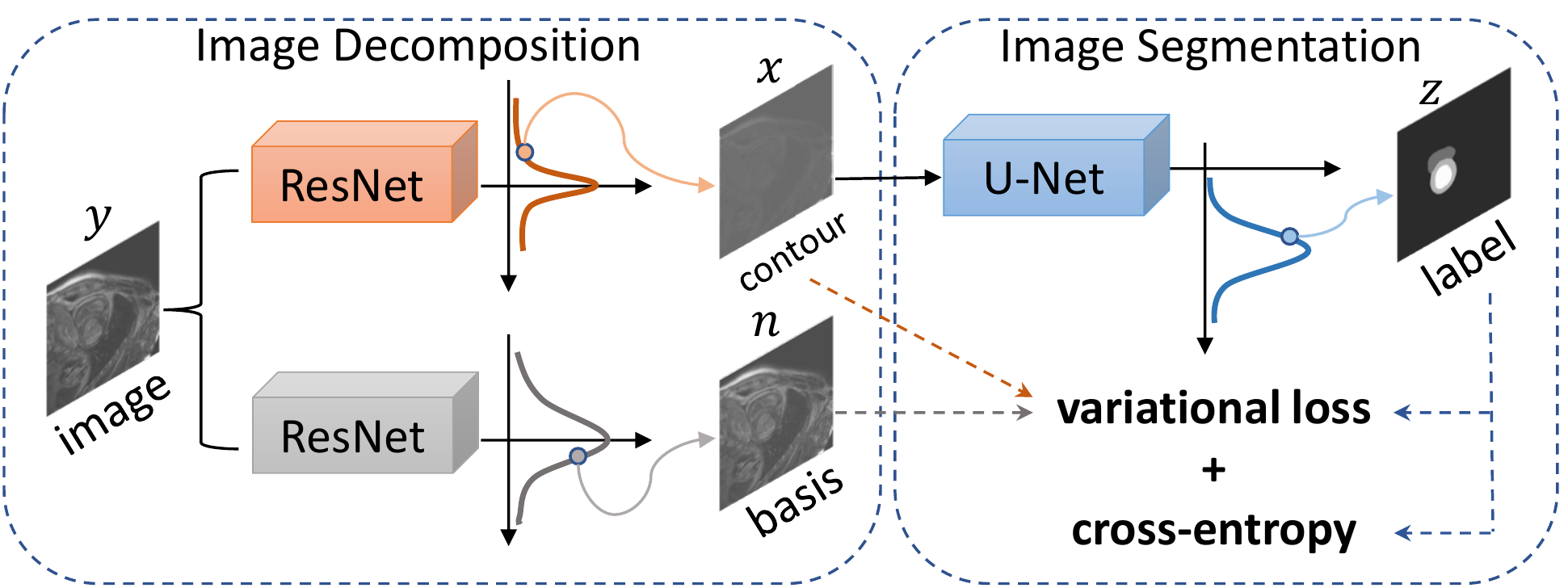}}
	\caption{The framework of Bayesian segmentation (BayeSeg). (a) shows the probabilistic graphical model (PGM) of BayeSeg. Here, the blue circle denotes an observed image, green and orange circles denote unknown variables, and white rectangles denote hyper-parameters. (b) presents the deep learning framework of BayeSeg. Given an image, we first use ResNets to infer the posterior distributions of the contour and basis, and obtain their random samples. Then, we use U-Net to infer the posterior distribution of label, and the resulting random sample is a segmentation. Please refer to Section \ref{sec:train} for the details of network architecture and training strategy.}
	\label{fig:framework}
\end{figure}

We propose a Bayesian segmentation (BayeSeg) framework to improve the generalizability of deep learning models. Many learning-based methods trained on one sequence MR images cannot generalize well to the other sequence or site data \cite{Li/2019}. To solve the challenge, we propose the BayeSeg mainly consisting of two parts, i.e., (1) statistical modeling of image and label as shown in Fig. \ref{fig:framework} (a), and (2) statistical inference of image and label as shown in Fig. \ref{fig:framework} (b). At the first stage, we build a probabilistic graphical model (PGM) for the modeling of image and label. That is we decompose an image into its contour and basis, and only the contour is related to an expected label. At the second stage, we first build two residual networks (ResNets) to infer the posterior distributions of the contour and basis, respectively. Then, we build a U-Net to estimate the posterior distribution of the label. An intuitive understanding of ``contour'' and ``basis'' is shape and appearance. Since shape is domain-irrelevant, the model predicting a label from the contour will have better generalizability.

Fig. \ref{fig:framework} shows the framework of statistical modeling and inference of the proposed BayeSeg. For the statistical modeling as shown in Fig. \ref{fig:framework} (a), we first decompose an image $\bm y$ into its basis $\bm n$ and contour $\bm x$. The former is a Gaussian variable depending on the mean $\bm m$ and the inverse variance $\bm \rho$. The latter depends on the expected label $\bm z$ and the line $\bm \upsilon$ for detecting the edges of contour. Similarly, the label $\bm z$ depends on the segmentation boundary $ \bm \omega $ and the segmentation probability $\bm \pi$ of all classes. Finally, Gamma priors are assigned to $\bm \rho$, $\bm \upsilon$, and $\bm \omega$, a Beta prior is assigned to $\bm \pi$, and a Gaussian prior is assigned to $\bm m$. Fig. \ref{fig:framework} (b) shows the deep learning framework of inferring related variables. The outputs of ResNets and U-Net will be jointly used to compute a variational loss, which is the key of improving model generalizability.   

\subsection{Statistical modeling of image and label}
This section shows the statistical modeling of image and label. Given an image sampled from the variable $\bm y \in \mathbb{R}^{d_{y}}$, where $d_{y}$ denotes the dimension of $\bm y$, we decompose $\bm y$ into the sum of a contour $\bm x$ and a basis $\bm n$, i.e., $\bm y = \bm x + \bm n$.
Then, the basis $\bm n$ is modeled by a normal distribution with a mean $\bm m \in \mathbb{R}^{d_{y}}$ and a covariance $diag(\bm \rho)^{-1} \in \mathbb{R}^{d_{y} \times d_{y}}$. Moreover, the contour $\bm x$ is modeled by a simultaneous autoregressive model (SAR) \cite{Shekhar/2008} depending on the expected label $\bm z \in \mathbb{R}^{d_{y} \times K} $, the line $\bm \upsilon \in \mathbb{R}^{d_{y}} $ indicating edges of the contour, and the matrix $\bm D_x \in \mathbb{R}^{d_{y} \times d_{y}} $ describing a neighboring system of the contour, where $K$ is the number of classes for segmentation. Finally, the observation likelihood of an image $\bm y$ can be expressed as
\begin{equation}\label{eq:likelihood}
	p(\bm y|\bm x,\bm m, \bm \rho) = \mathcal{N}(\bm x + \bm m, diag(\bm \rho)^{-1}).
\end{equation}

The basis $\bm n$ of an image is determined by a normal distribution, that is, $ p(\bm n|\bm m,\bm \rho) = \mathcal{N}(\bm n|\bm m, diag(\bm \rho)^{-1}) $. Specifically, we assign a Gaussian prior to $\bm m$, i.e., $p(\bm m|\bm \mu_{0}, \sigma_{0}) = \mathcal{N}(\bm m|\bm \mu_{0}, \sigma_{0}^{-1}\bm I)$, and a Gamma prior to $\bm \rho$, namely, $p(\bm \rho|\bm \phi_{\rho},\bm \gamma_{\rho}) = \prod^{d_{y}}_{i=1}\mathcal{G}(\rho_{i}|\phi_{\rho i}, \gamma_{\rho i})$. Here, $\bm I$ denotes an identity matrix, $\mu_{0}, \sigma_{0}, \phi_{\rho}$, and $\gamma_{\rho}$ are predefined hyper-parameters, and $\mathcal{G}(\cdot,\cdot)$ represents the Gamma distribution.

The contour $\bm x$ of an image is determined by a SAR mainly depending on the expected label $\bm z$ and the line $\bm \upsilon$,
\begin{equation}\label{eq:contour}
	p(\bm x|\bm z, \bm \upsilon)
	= \textstyle\prod_{k=1}^{K} \mathcal{N}(\bm x|\bm 0, [\bm D_{x}^{T}diag(\bm z_{k}\bm\upsilon)\bm D_{x}]^{-1}),
\end{equation}
where, $\bm z_{k}$ denotes the segmentation of the $k$-th class, and $\bm D_x = \bm I - \bm B_{x}$ is non-singular. $\bm B_{x}$ describes a neighboring system of each pixel. For examples, if the values of $\bm B_{x}$ for the nearest four pixels equals to 0.25 while others are zeros, then $ \bm D_x $ is aimed to compute the average difference of each pixel with its four neighbors. The line $\bm \upsilon$ can indicate the edges of the contour, and it is assigned a Gamma prior, i.e., $p(\bm\upsilon|\phi_{\upsilon i}, \gamma_{\upsilon i}) = \prod_{i=1}^{d_{y}}\mathcal{G}(\upsilon_{i}|\phi_{\upsilon i}, \gamma_{\upsilon i})$.

The label $\bm z$ is modeled by another SAR depending on the segmentation boundary $\bm \omega \in \mathbb{R}^{d_{y} \times K}$ and the segmentation probability $\bm \pi \in \mathbb{R}^K$ of all classes, namely,
\begin{equation}\label{eq:label}
	p(\bm z|\bm \pi, \bm \omega) = \textstyle\prod_{k=1}^{K} \mathcal{N}(\bm z|\bm 0, [-\ln (1-\pi_{k})\bm D_{z}^{T}diag(\bm \omega_{k})\bm D_{z}]^{-1}),
\end{equation}
where, the definition of $\bm D_z$ is the same as $\bm D_x$ in Eq. (\ref{eq:contour}); $\bm \omega_k$ can indicate the boundary of the $k$-th segmentation $\bm z_k$; and $\pi_k$ denotes the probability of a pixel belonging to the $k$-th class. Finally, we assign Gamma prior to $\bm \omega$, i.e., $p(\bm\omega) = \prod_{i=1}^{d_{y}}\prod_{k=1}^{K}\mathcal{G}(\omega_{ki}|\phi_{\omega ki}, \gamma_{\omega ki})$, and give Beta prior to $\bm\pi$, namely, $p(\bm\pi) = \prod_{k=1}^{K}\mathcal{B}(\pi_{k}|\alpha_{\pi k}, \beta_{\pi k})$. \emph{The details of Gaussian distribution, Gamma distribution, and Beta distribution are provided in Appendix \ref{appendA}.}

\subsection{Variational inference of image and label}
This section shows a variational method of inferring the contour, basis and label given an image $\bm y$ by maximum a \textit{posteriori} (MAP) estimation. Let $\bm\psi=\{\bm m, \bm\rho, \bm x, \bm\upsilon, \bm z, \bm\omega, \bm\pi\}$ denote the set of all variables to infer, then our aim is to infer the posterior distribution $p(\bm\psi|\bm y)$. Since direct computation is intractable, we use variational Bayesian (VB) method \cite{Blei/2017} to solve the problem. Concretely, we approximate the posterior distribution $p(\bm\psi|\bm y)$ via a variational distribution $q(\bm\psi)$ by assuming the variables in $\bm\psi$ are independent, namely,
\begin{equation}\label{eq:variational}
	q(\bm\psi) = q(\bm m)q(\bm\rho)q(\bm x)q(\bm\upsilon)q(\bm z)q(\bm\omega)q(\bm\pi).
\end{equation}
After that, we minimize the KL divergence between $q(\bm\psi)$ and $p(\bm\psi|\bm y)$, and which results in our final variational loss as follows,
\begin{equation}\label{eq:varloss}
	\mathop{\min}_{q(\bm\psi)} \mathcal L_{var} = \mbox{KL}(q(\bm\psi)||p(\bm\psi)) - \mathbb{E}[\ln p(\bm y| \bm\psi)]
\end{equation}
\emph{The details of further unfolding the variational loss are provided in Appendix \ref{appendB}.} 

\subsection{Neural networks and training strategy}\label{sec:train}
This section shows the network architecture of achieving the variational inference and the training strategy for image segmentation. As Fig. \ref{fig:framework} (b) shows, \emph{at the decomposition stage}, we adopt two ResNets \cite{He/2016} to separately infer the variational posteriors of the contour $\bm x$ and basis $\bm n$, i.e., $q(\bm x)$ and $q(\bm n)$, respectively. The ResNet of inferring the contour consists of 10 residual blocks, and each block has a structure of ``Conv + ReLU + Conv''. The output of this ResNet has two channels. One is the element-wise mean of the contour, and the other is its element-wise variance. The contour $\bm x$ in the figure denotes a random sample from $q(\bm x)$. The ResNet of inferring the basis consists of 6 residual blocks, and each block has a structure of ``Conv + BN + ReLU + Conv + BN''. Similarly, this ResNet will output the mean and variance of the basis, and the basis $\bm n$ is randomly sampled from its variational posterior distribution. \emph{At the segmentation stage}, we adopt a U-Net \cite{unet/2015} to infer the variational posterior of the label $\bm z$, i.e., $q(\bm z)$. The output of this U-Net has $2K$ channels. The first $K$ channels denote the element-wise mean of the label, and the left channels represent its element-wise variance. The label $\bm z$ in Fig. \ref{fig:framework} (b) is a random sample from the resulting posterior distribution, and it will be taken as a stochastic segmentation for training.

BayeSeg is trained in an end-to-end manner by balancing between cross-entropy and the variational loss in (\ref{eq:varloss}). For convenience, the cross-entropy between a stochastic segmentation and the provided manual segmentation is notated as $\mathcal L_{ce}$. Then, our total loss of training BayeSeg is given by,
\begin{equation}\label{eq:totalloss}
	\min_{q(\psi)} \mathcal L_{ce} + \lambda \mathcal L_{var},
\end{equation}
where, the balancing weight $\lambda$ is set to 100 in our experiments. Besides, other hyper-parameters in Fig. \ref{fig:framework} (a) is summarized as follows. Each element of $\bm \phi_{\cdot}$ of related variables is set to 2; $\alpha_{\pi}=2$ and $\beta_{\pi}=2$ for the segmentation probability $\bm \pi$; $\bm \mu_0 = \bm 0$ and $\sigma_0 = 1$ for the mean of the basis; The elements of $\gamma_{\rho}$, $\gamma_{\upsilon}$, and $\gamma_{\omega}$ are set to $10^{-6}$, $10^{-8}$, and $10^{-4}$, respectively. Note that BayeSeg could properly decompose an image into the contour and basis due to the priors with respect to $ \bm \upsilon $ and $ \bm \rho $. That is the contour and basis are adaptively balanced after selecting proper $\gamma_{\rho}$ and $\gamma_{\upsilon}$. 

\section{Experiments}
\subsection{Tasks and datasets}
We used the LGE of MSCMRseg \cite{zhuang/2019} to train models. To validate the performance on the task of cross-sequence segmentation, we tested models using LGE and T2 of MSCMRseg. To validate the performance on the task of cross-site segmentation, we tested models using LGE of MSCMRseg and ACDC \cite{ACDC/2018}. 

MSCMRseg \cite{zhuang/2016,zhuang/2019} was provided by MICCAI'19 Multi-sequence Cardiac MR Segmentation Challenge. This dataset consists of 45 multi-sequence CMR images, including LGE, C0, and T2. Each case comes from the same patient who underwent cardiomyopathy. The manual segmentation results of left ventricle (LV), right ventricle (RV), and myocardium (Myo) for all images are available. In this study, we randomly split the 45 cases into three sets consisting of 25, 5, and 15 cases, respectively. Then, we only used the 25 LGE images for training, and the 5 LGE images for validation. Finally, we tested models on the 15 multi-sequence cases to show the performance of cross-sequence segmentation.

ACDC \cite{ACDC/2018} was provided by MICCAI'17 Automatic Cardiac Diagnosis Challenge. This dataset consists of shot-axis cardiac cine-MRIs of 100 patients for training, and of 50 patients for test. Only the manual segmentation results of training data are provided for LV, RV, and Myo during the end-diastolic (ED) and end-systolic (ES) phases. In our study, we tested models using the 100 training images to show the performance of cross-site segmentation.

BayeSeg was implemented by Pytorch, and trained by Adam optimizer with the initial learning rate to be $10^{-4}$. The learning rate was dropped every 500 epochs by a factor of 0.1, and the training was stopped when up to 2000 epochs. At the test stage, we took the mean of $ \mathbf z $ as the final segmentation label, since the variational posterior $ q(\mathbf z) $ is a Gaussian distribution whose mode is its mean. All experiments were run on a TITAN RTX GPU with 24G memory.

\subsection{Cross-sequence segmentation}
To study the performance of BayeSeg on the task cross-sequence segmentation, we trained four models using the 25 LGE images of MSCMRseg. Concretely, we trained a U-Net, which had the same architecture as the U-Net in Fig. \ref{fig:framework} (b), by minimizing the cross-entropy. Then, we trained PU-Net on the same dataset using its public code and default settings. Moreover, we trained a baseline, which shares the same architecture as BayeSeg, without using the variational loss in (\ref{eq:totalloss}). Finally, we trained the BayeSeg by minimizing the total loss in (\ref{eq:totalloss}). For fair comparisons, all models were trained using consistent data augmentation, including random flip and rotation. In the test stage, we evaluated all models on the 15 multi-sequence images, i.e., LGE and T2, and reported the dices of LV, Myo, and RV as well as average dice.

\begin{figure}[t]
	\centering
	\includegraphics[width=0.95\linewidth]{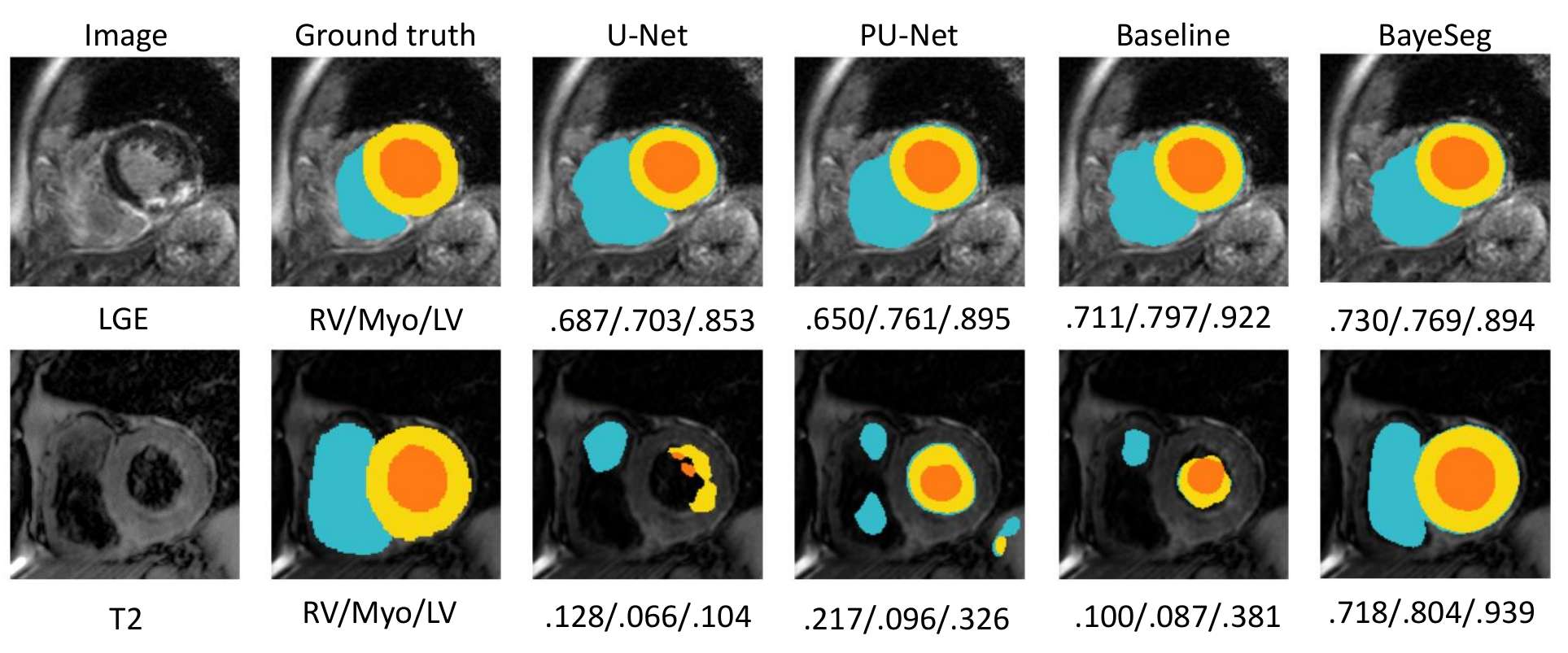}\\[-3ex]
	\caption{Visualization of results on the task of cross-sequence segmentation. Here, we choose the median case of BayeSeg according to the dices of 15 LGE images.}
	\label{fig:cross:sequence}
\end{figure}

Table \ref{tab:cross:sequence} reports the results of compared methods. One can see that PU-Net, Baseline, and BayeSeg achieved comparable performance, when the training and test sequences were consistent. If the training sequence and test sequence are different, the performance of BayeSeg dropped weakly, but that of others decreased dramatically. Besides, BayeSeg greatly outperformed Baseline in this case, which demonstrates that the variational loss, induced by joint modeling of image and label statistics, is the key of improving model generalizability. To qualitatively evaluate all models, we chose the median case of BayeSeg according to the average dices of 15 LGE images, and visualized the segmentation results in Figure~\ref{fig:cross:sequence}. This figure shows BayeSeg delivers the best performance in segmenting the unseen T2 sequence, which again confirms the effectiveness of our framework in improving model generalizability.

\begin{table}[!t]
	\centering
	\caption{Evaluation on the task of cross-sequence cardiac segmentation. \emph{Note that all models were only trained on LGE of MSCMRseg, but tested on LGE and T2}. Here, G denotes the drop of average dice, and it is used to measure model generalizability.}
	\resizebox{1\linewidth}{!}{
		\begin{tabular}{|c|cccc|cccc|c|}
			\hline
			\multirow{2}{*}{Method}& \multicolumn{4}{c|}{LGE of MSCMRseg (15 samples)}& \multicolumn{4}{c|}{T2 of MSCMRseg (15 samples)}& \multirow{2}{*}{G}\\
			\cline{2-9}
			&  LV&  Myo&  RV&  Avg&  LV&  Myo&  RV&  Avg&  \\
			\hline
			U-Net&  .855$\pm$.045&  .727$\pm$.064&  .733$\pm$.097&  .772$\pm$.093&  .203$\pm$.183&  .095$\pm$.093&  .055$\pm$.062&  .118$\pm$.139&  .654\\
			\hline
			PU-Net&  \textbf{.898$\pm$.027}&  .768$\pm$.056&  .729$\pm$.089&  .798$\pm$.096&  .279$\pm$.162&  .166$\pm$.122&  .195$\pm$.130&  .213$\pm$.147&  .585\\
			\hline
			Baseline&  .893$\pm$.023&  \textbf{.783$\pm$.045}&  .727$\pm$.069&  .801$\pm$.085& .481$\pm$.129&  .117$\pm$.079&  .090$\pm$.123&  .230$\pm$.211&  .571\\
			\hline
			BayeSeg&  .887$\pm$.028&  .774$\pm$.048&  \textbf{.763$\pm$.060}&  \textbf{.808$\pm$.073}& \textbf{.846$\pm$.119}&  \textbf{.731$\pm$.117}&  \textbf{.528$\pm$.206}&  \textbf{.701$\pm$.202}&  \textbf{.107}\\
			\hline
	\end{tabular}}
	\label{tab:cross:sequence}
\end{table}

\subsection{Cross-site segmentation}
To study the performance of BayeSeg on the task of cross-center segmentation, we tested all models trained in the previous section on ACDC. Table~\ref{tab:cross:center} reports the results of these models. This table showed that the performance of all methods dropped when the training and test samples came from two different sites, but BayeSeg delivered the least performance drop. Therefore, BayeSeg generalized well on the unseen samples from the different site, thanks to the joint modeling of image and label statistics. 

\begin{table}[!t]
	\centering
	\caption{Evaluation on the task of cross-site cardiac segmentation. \emph{Note that all models were only trained on LGE of MSCMRseg, but tested on LGE and ACDC}. Here, G denotes the drop of average dice, and it is used to measure model generalizability.}
	\resizebox{1\linewidth}{!}{
		\begin{tabular}{|c|cccc|cccc|c|}
			\hline
			\multirow{2}{*}{Method}& \multicolumn{4}{c|}{LGE of MSCMRseg (15 samples)}& \multicolumn{4}{c|}{ACDC (100 samples)}& \multirow{2}{*}{G}\\
			\cline{2-9}
			&  LV&  Myo&  RV&  Avg&  LV&  Myo&  RV&  Avg&  \\
			\hline
			U-Net&  .855$\pm$.045&  .727$\pm$.064&  .733$\pm$.097&  .772$\pm$.093&  .721$\pm$.187&  .602$\pm$.183&  .659$\pm$.202&  .660$\pm$.197&  .112\\
			\hline
			PU-Net&  \textbf{.898$\pm$.027}&  .768$\pm$.056&  .729$\pm$.089&  .798$\pm$.096&  .743$\pm$.152&  .641$\pm$.146&  .604$\pm$.215&  .663$\pm$.184&  .126\\
			\hline
			Baseline&  .893$\pm$.023&  \textbf{.783}$\pm$.045&  .727$\pm$.069&  .801$\pm$.085&  .776$\pm$.134&  .667$\pm$.150&  .585$\pm$.227&  .676$\pm$.192&  .125\\
			\hline
			BayeSeg&  .887$\pm$.028&  .774$\pm$.048&  \textbf{.763$\pm$.060}&  \textbf{.808$\pm$.073}&  \textbf{.792$\pm$.130}&  \textbf{.694$\pm$.123}&  \textbf{.659$\pm$.175}&  \textbf{.715$\pm$.155}&  \textbf{.093}\\
			\hline
	\end{tabular}}
	\label{tab:cross:center}
\end{table}

% \begin{figure}[htp]
	%     \centering
	%     \includegraphics[width=1\linewidth]{figs/crossSite.png}
	%     \caption{Visualization of results on the task of cross-site segmentation.}
	%     \label{fig:cross:site}
	% \end{figure}
\subsection{Interpretation of joint modeling}
\begin{figure}[t]
	\centering
	\includegraphics[width=0.9\linewidth]{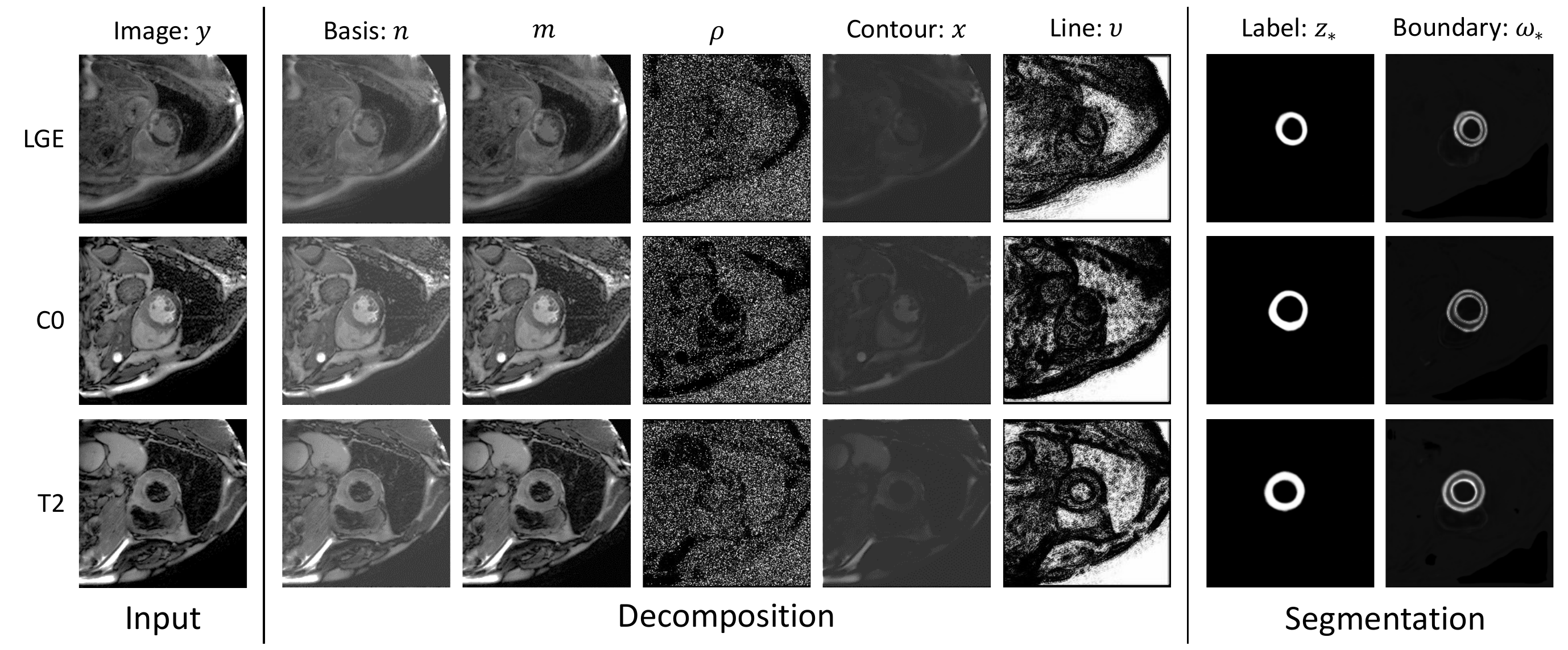}\\[-2.5ex]
	\caption{Visualization of posteriors inferred by BayeSeg. Here, the subscript $_*$ denotes the result of myocardium.}
	\label{fig:posteriors}
\end{figure}

In this section we interpreted the joint modeling of image and label statistics. Fig. \ref{fig:posteriors} shows the posteriors inferred by our BayeSeg for three different sequences of MSCMRseg. One can see that, at the decomposition stage, an image was mainly decomposed into its basis and contour. The basis $\bm n$ was modeled as a Gaussian distribution with the mean $\bm m$ and the inverse variance $\bm \rho$. It was an approximation of the image, and therefore $\bm x$ was left as the contour. To avoid the smoothness of this contour, we assigned the line $\bm \upsilon$ to detect its edges. The large values of $\bm \upsilon$ indeed showed the smooth areas of contour, while the small values indicate the edges. At the segmentation stage, we choose to segment the contour, since it is more likely to be sequence-independent, site-independent, and even modality-independent. To achieve better segmentation around the boundary of some object, such as myocardium, we assigned the $\omega$ to detect the segmentation boundary. This variable successfully indicated the inner and outer boundaries of myocardium, as shown in Fig. \ref{fig:posteriors}.

\section{Conclusion}
In this work, we proposed a new Bayesian segmentation framework by joint modeling of image and label statistics. Concretely, we decomposed an image into its basis and contour, and estimated the segmentation of this image from the more stable contour. Our experiments have shown that the proposed framework could address the problem of over-fitting and greatly improve the generalizability of deep learning models. 

\appendix

\section{Preliminary}\label{appendA}
If $ n $ is a variable which follows Gaussian distribution, then its probability density function is given by 
\begin{equation}\label{Gauss}
	p(n|m, \rho)= \mathcal N(n|m, \rho^{-1}) = \frac{1}{\sqrt{2\pi/\rho}}\exp^{-\frac{\rho}{2}(n-m)^2},
\end{equation}

If $ \omega $ is a variable which follows Gamma distribution, then its probability density function is given by
\begin{equation}\label{Gamma}
	p(\omega|\phi,\gamma)= \mathcal G(\omega|\phi, \gamma) = \frac{\phi^\gamma}{\Gamma(\gamma)} \omega^{\gamma - 1} e^{-\phi \omega},
\end{equation}
where, $ \Gamma(\cdot) $ denotes the Gamma function.

If $\pi$ is a variable which follows Beta distribution, then its probability density function is given by
\begin{equation}\label{Beta}
	p(\pi|\alpha,\beta)= \mathcal B(\omega|\alpha, \beta) = \frac{\Gamma(\alpha + \beta)}{\Gamma(\alpha) \Gamma(\beta)} \pi^{\alpha - 1} (1-\pi)^{\beta - 1}.
\end{equation}

\section{Variational Inference}\label{appendB}
To estimate the variational posteriors, we minimize the KL divergence between $q(\bm\psi)$ and $p(\bm\psi|\bm y)$, which results in
\begin{equation}
	\mathop{\arg\!\min}_{q(\bm\psi)} \mbox{KL}(q(\bm\psi)||p(\bm\psi|\bm y)) = \mathop{\arg\!\min}_{q(\bm\psi)} \mbox{KL}(q(\bm\psi)||p(\bm\psi)) - \mathbb{E}[\ln p(\bm y| \bm\psi)], \tag{1} \label{1}
\end{equation}
Moreover, we covert it to the following problem by reparameterization,
\begin{equation}
	\mathop{\arg\!\min}_{q(\bm\psi)} \mbox{KL}(q(\bm\psi)||p(\bm\psi)) -\mathbb{E}_{q(\bm\rho)}[\ln p(\bm y|\bm x, \bm m, \bm\rho)]. \label{2}\tag{2}
\end{equation}

\subsection{Explicit computation of  $q(\bm \upsilon),q(\bm \omega),q(\bm \pi),q(\bm \rho)$}
Minimizing (\ref{2}) over $q(\bm \upsilon),q(\bm \omega),q(\bm \pi),q(\bm \rho)$ successively results in the explicit formulas of computing the parameters of these distributions as follows,
\begin{equation*}
	\left\{
	\begin{aligned}
		&\hat{\alpha}_{\upsilon i} = \gamma_{\upsilon i} + K/2 \\
		&\hat{\beta}_{\upsilon i} = \frac{1}{2}\textstyle\sum_{k=1}^{K}\hat{\bm\mu}_{zki}[(\bm D_{x}\hat{\bm\mu}_{x})^{2}_{i} + \langle\hat{\bm\sigma}_{x},\bm d_{xi}^{2}\rangle] + \phi_{\upsilon i} \\
		&\hat{\bm\mu}_{\upsilon} = \frac{\hat{\bm\alpha}_{\upsilon}}{\hat{\bm\beta}_{\upsilon}} =\frac{2\bm\gamma_{\upsilon} + K}{\sum_{k=1}^{K}\hat{\bm\mu}_{zk}[(\bm D_{x}\hat{\bm\mu}_{x})^{2} + 2\hat{\bm\sigma}_{x}] + 2\bm\phi_{\upsilon}}
	\end{aligned}
	\right.,
\end{equation*}
\begin{equation*}
	\left\{
	\begin{aligned}
		&\hat{\alpha}_{\omega ki} = \gamma_{\omega ki} + 1/2 \\
		&\hat{\beta}_{\omega ki} = \frac{1}{2}[ \Psi(\hat{\alpha}_{\pi k}+\hat{\beta}_{\pi k}) - \Psi(\hat{\beta}_{\pi k})][(\bm D_{z}\hat{\bm\mu}_{zk})_{i} + \langle\hat{\bm\sigma}_{zk}^{2}, \bm d_{zi}^{2}\rangle] + \phi_{\omega ki} \\
		&\hat{\bm\mu}_{\omega k} = \frac{\hat{\bm\alpha}_{\omega k}}{\hat{\bm\beta}_{\omega k}} =\frac{2\bm\gamma_{\omega k} + 1}{[ \Psi(\hat{\alpha}_{\pi k}+\hat{\beta}_{\pi k}) - \Psi(\hat{\beta}_{\pi k})][(\bm D_{z}\hat{\bm\mu}_{zk})^{2} + 2{\bm\sigma}_{zk}^{2}] + 2\phi_{\omega k}}
	\end{aligned}
	\right.,
\end{equation*}
\begin{equation*}
	\left\{
	\begin{aligned}
		&\hat{\alpha}_{\pi k} = \alpha_{\pi k} + d_y/2 \\
		&\hat{\beta}_{\pi k} = \frac{1}{2} \textstyle\sum_{i=1}^{d_y} \hat{\bm\mu}_{\omega k i} [(\bm D_{z}\hat{\bm\mu}_{zk})_i^{2} + 2{\bm\sigma}_{zki}^{2}] + \beta_{\pi k} \\
	\end{aligned}
	\right.,
\end{equation*}
and $ \hat{\bm\mu}_{\rho} = \hat{\bm\alpha}_{\rho}/\hat{\bm\beta}_{\rho} = (2\gamma_{\rho}+1)/([\bm y - (\bm x + \bm m)]^{2} + 2\phi_{\rho}) $. Here, $ \Psi(\cdot) $ denotes the Digamma function. Finally, the related variational posterior distributions are given by
\begin{align*}
	&q(\bm\upsilon) = \prod^{d_{y}}_{i=1}\mathcal{G}(\upsilon_{i}|\hat{\beta}_{\upsilon i}, \hat{\alpha}_{\upsilon i}) \mbox{ and } q(\bm\omega) = \prod_{k=1}^{K}\prod_{i=1}^{d_{y}}\mathcal{G}(\omega_{ki}|\hat{\beta}_{\omega ki}|\hat{\beta}_{\omega ki}, \hat{\alpha}_{\omega ki})\\
	& q(\bm\pi) = \prod_{k=1}^{K}\mathcal{B}(\pi_{k}|\hat{\alpha}_{\pi k}, \hat{\beta}_{\pi k}) \mbox{ and } q(\bm \rho) = \prod^{d_{y}}_{i=1}\mathcal{G}(\rho_{i}|\hat{\beta}_{\rho i}, \hat{\alpha}_{\rho i})
\end{align*}

\subsection{Variational inference of $q(\bm x), q(\bm z)$ and $q(\bm m)$}
Minimizing (\ref{2}) over $q(\bm x), q(\bm z)$ and $q(\bm m)$ successively results in the losses of further inferring the parameters of these distributions as follows,  
\begin{equation*}
	\mathcal{L}_{y} = \frac{1}{2}||\bm y - (\bm x + \bm m)||^{2}_{diag(\hat{\bm\mu}_{\rho})},
\end{equation*}
where, $ \bm x = \hat{\bm\sigma}_x \odot \bm \epsilon + \hat{\bm\mu}_x $, $ \bm m = \hat{\bm\sigma}_m \odot \bm \epsilon + \hat{\bm\mu}_m $, and $ \bm \epsilon \sim \mathcal N(\bm 0, \bm I) $.
\begin{gather*}
	\left\{
	\begin{aligned}
		&\mathcal{L}_{\hat{\mu}_{z}} = \frac{1}{2}\textstyle\sum_{k=1}^{K}[\Psi(\hat{\alpha}_{\pi k}+\hat{\beta}_{\pi k}) - \Psi(\hat{\beta}_{\pi k})]||\bm D_{z}\hat{\bm\mu}_{zk}||^{2}_{diag(\hat{\bm\mu}_{\omega k})} \\
		&\mathcal{L}_{\hat{\sigma}_{z}} = \frac{1}{2}\textstyle\sum_{k=1}^{K}[\Psi(\hat{\alpha}_{\pi k}+\hat{\beta}_{\pi k}) - \Psi(\hat{\beta}_{\pi k})]\left[ \langle 2\hat{\bm\mu}_{\omega k}, \hat{\bm\sigma}_{zk}^{2}\rangle - \langle\bm 1, \ln(\hat{\bm\sigma}_{zk}^{2})\rangle\right]
	\end{aligned}
	\right.
\end{gather*}
\begin{gather*}
	\left\{
	\begin{aligned}
		&\mathcal{L}_{\hat{\mu}_{x}} = \frac{1}{2}\textstyle\sum_{k=1}^{K}||\bm D_{x}\hat{\bm\mu}_{x}||^{2}_{diag(\hat{\bm\mu}_{zk}\hat{\bm\mu}_{\upsilon})} \\
		&\mathcal{L}_{\hat{\sigma}_{x}} = \frac{1}{2}\textstyle\sum_{k=1}^{K}\left[\langle 2\hat{\bm\mu}_{zk}\hat{\bm\mu}_{\upsilon},\hat{\sigma}_{x}^{2}\rangle - \frac{1}{K}\langle\bm 1, \ln(\hat{\bm\sigma}_{x}^{2})\rangle\right]
	\end{aligned}
	\right.
\end{gather*}
\begin{gather*}
	\left\{
	\begin{aligned}
		&\mathcal{L}_{\hat{\mu}_{m}} = \frac{\sigma_{0}}{2}||\hat{\bm\mu}_{m}||^{2}_{2} \\
		&\mathcal{L}_{\hat{\sigma}_{m}} = \frac{1}{2}[\langle\sigma_{0}\bm 1, \hat{\bm\sigma}_{m}^{2}\rangle - \langle\bm 1, \ln(\hat{\bm\sigma}_{m}^{2})]
	\end{aligned}
	\right.
\end{gather*}
Overall, the final \textbf{variational loss} of inferring the basis, contour, and label of an image is summarized as
\begin{equation*}
	\mathcal L_{var} = \mathcal{L}_{y} + \mathcal{L}_{\hat{\mu}_{z}} + \mathcal{L}_{\hat{\sigma}_{z}} + \mathcal{L}_{\hat{\mu}_{x}} + \mathcal{L}_{\hat{\sigma}_{x}} + \mathcal{L}_{\hat{\mu}_{m}} + \mathcal{L}_{\hat{\sigma}_{m}}.
\end{equation*}
Finally, the related variational posterior distributions are given by
\begin{align*}
	q(\bm m) &= \mathcal{N}(\bm m| \hat{\bm\mu}_{m}, diag(\hat{\bm\sigma}_{m}^{2})), \\
	q(\bm x) &= \mathcal{N}(\bm x| \hat{\bm \mu}_{x}, diag(\hat{\bm\sigma}_{x}^{2})), \\ 
	q(\bm z) &= \prod_{k=1}^{K}\mathcal{N}(\bm z| \hat{\bm \mu}_{zk}, diag(\hat{\bm\sigma}_{zk}^{2})).
\end{align*}

\end{document}